\documentclass[12pt, bibother, otherbib]{article}
\usepackage{isabelle,isabellesym}
\usepackage[utf8]{inputenc} 
\usepackage[T1]{fontenc}    
\fontencoding{T1}\selectfont
\usepackage{hyperref}       
\usepackage{url}            
\usepackage{booktabs}       
\usepackage{amsfonts}       
\usepackage{nicefrac}       
\usepackage{lipsum}
\usepackage{amsmath}
\usepackage{amsfonts}
\usepackage{amssymb}
\usepackage{graphicx}
\usepackage{hyperref}

\usepackage{amsmath,amssymb}
\usepackage{xpatch}

\let\ts=\thinspace

\newtheorem{Theorem}{Theorem}[section]
\newtheorem{theorem}{Theorem}[section]

\newtheorem{corollary}[Theorem]{Corollary}

\title{Irrationality and Transcendence Criteria for Infinite Series in Isabelle/HOL}

\begin{document}
\author{
Angeliki Koutsoukou-Argyraki 
 (\texttt{ak2110@cam.ac.uk})
 \and
Wenda Li
 (\texttt{wl302@cam.ac.uk})
 \and
 Lawrence C. Paulson FRS (\texttt{lp15@cam.ac.uk})\\ \\
Computer Laboratory, University of Cambridge\\
   15 JJ Thomson Avenue, Cambridge CB3 0FD, UK}

\maketitle

\begin{abstract}We give an overview of our formalizations in the proof assistant Isabelle/HOL of certain irrationality and transcendence criteria for infinite series from 
three different research papers: by Erd\H{o}s and Straus (1974), Han\v{c}l (2002) and Han\v{c}l and Rucki (2005). 
Our formalizations in Isabelle/HOL can be found on the Archive of Formal Proofs. Here we describe selected aspects of the formalization and  discuss what this reveals about 
the use and potential of Isabelle/HOL in formalizing modern mathematical research, particularly in these parts of number theory and analysis.
\end{abstract}

\emph{AMS 2020 subject classes}: 40A05, 11J68, 11J81, 03B35, 68V20, 68V35.

\section{Introduction}
The libraries of formalized mathematics in various proof assistants (or interactive theorem provers) have been growing for years.
However, the field is still far from the goal propounded in 1994 as the QED Manifesto~\cite{qed_manifesto} ``to build a computer system that effectively represents all important mathematical knowledge and techniques''.
We are light years from the vision of 
having an interactive tool that could ``converse'' with human mathematicians to aid in the discovery of new results, as  
colorfully described by Timothy Gowers~\cite{gowers}.  
Our goal within the ALEXANDRIA Project at Cambridge~\cite{alexandria}
is to contribute to making proof assistants (and more specifically Isabelle/HOL) more useful to research mathematicians.
To gather information about what specifically needs to be done, we have been formalizing material to contribute to the Isabelle/HOL Libraries\footnote{\url{isabelle.in.tum.de/dist/library/HOL}} and the Archive of Formal Proofs (AFP).\footnote{\url{www.isa-afp.org}}
As research mathematics is our main target, and given that a considerable amount of basic mathematics has already been formalized in Isabelle/HOL over the years, here we wish to explore
the following \textbf{question}:
\begin{quote}\em
How receptive to formalization (particularly in Isabelle/HOL) is mainstream published mathematical research? 
    \end{quote}
In this paper we are not working towards the ambitious goal of formalizing the proof of a landmark result.
Nor are we concerned with standard material from undergraduate or graduate textbooks. Those proofs have been thoroughly studied and checked, 
and repeatedly reworked and polished by many people over the years.
Instead, we are formalizing mainstream mathematical research from journal papers in their raw form, not reworked or clarified as textbook proofs normally are.
To explore the question above (within a specific case study), we ask questions such as these:
\begin{itemize}
\item To what extent does the current Isabelle/HOL library cover the necessary preliminaries?
\item How helpful are the automatic proof tools?
\item  How much longer than the original proofs are the formalized proofs  (i.e.\ the \emph{de Bruijn factor} \cite{freekdebruijn})?
\item Are we going to discover any issues or gaps in the proofs?
\end{itemize}
Clearly, the answers to these questions depend on the chosen proofs. Here we attempt some case studies 
in number theory, on transcendence and irrationality criteria for infinite series.
In particular, we choose several results from three research papers by Erd\H{o}s and Straus (1974) \cite{erdosstraus}, Han\v{c}l (2002)~\cite{hancl}, and Han\v{c}l and Rucki (2005)~\cite{hanclrucki}.
This line of research makes heavy use of standard background material from analysis, e.g.\ convergence tests for series and (for Erd\H{o}s and Straus \cite{erdosstraus}) certain results about prime numbers. The material from Han\v{c}l and Rucki~\cite{hanclrucki} applies Roth's theorem on rational approximations to algebraic numbers.\footnote{The proof of Roth's theorem itself has not been formalized; see Section 5.}
Inevitably, we had to pick topics in which we a priori knew that at least some of the preliminaries were already available in Isabelle/HOL\@.
The material we chose is mathematically interesting in its own right, and moreover---unlike most results formalized nowadays---it comes
from fairly recent research papers instead of textbooks. 

Paul Erd\H{o}s was keenly interested in establishing criteria for the irrationality and transcendence of infinite series \cite{erdosenseignement,erdosliouville, erdosstrausnumbertheoretic}.
Several techniques for showing the transcendence of series are given by Nishioka \cite{nishioka}, who presents, among other,
methods originated by Mahler \cite{mahler}.
Later, S\'{a}ndor \cite{sandor}, Nyblom \cite{nyblom2000, nyblom2001}, and Han\v{c}l \cite{Hanclslov, hancltwo, Hanclcriterion} obtained many more results in this area. 

In this paper, we merely present in standard mathematical language the statements (without the informal proofs) of the theorems that we formalized and afterwords we discuss our formalization in Isabelle/HOL focussing on its most interesting aspects. 
For the informal proofs the reader may refer to the respective papers \cite{erdosstraus, hancl, hanclrucki}. Our full 
Isabelle/HOL formalizations can be found online, on the AFP \cite{irrationalityerdosstraus, irrationalityhancl, transhanclrucki}. 
As we had to fill in many intermediate arguments that were implicit in the original proofs, providing more details and clarifications, 
our formalizations may serve not only as case studies to explore the question of how receptive to formalization in Isabelle/HOL mathematical research is and as proof verification, but also as helpful supplementary material for a detailed study of the original proofs.

The plan of this paper is as follows: in the next section we give a brief introduction to Isabelle/HOL\@. In Sections 3, 4 and 5 we present the material formalized 
and discuss the main challenges and lessons learned through our formalizations for each paper respectively.
In Section 6 we briefly discuss the de Bruijn factors of the formalizations and finally, in Section 7 we present our conclusion.

\section{Isabelle/HOL}
Isabelle\footnote{\url{https://isabelle.in.tum.de/}} is an interactive theorem prover first developed in the 1980s by Lawrence Paulson and Tobias Nipkow \cite{isabelle_book, paulson89}. Today, Isabelle provides the \textit{Isar} language for writing proofs. Isar proofs are hierarchically structured, and they include enough redundancy---consisting of explicitly stated assumptions and goals---to be understandable to humans as well as machines. Isabelle supports multiple different logical formalisms, with Isabelle/HOL specifically based on higher order logic. 
Unlike proof assistants based on constructive type theories with dependent types, it implements simple type theory.
Isabelle/HOL admits classical (non-constructive) proofs, i.e.\ it accepts the law of the excluded middle, $P\lor\neg P$, as well as the axiom of choice.

Simple type theory and classical logic allow for powerful automation, which is a significant advantage.
Isabelle/HOL's main automation tool is \textit{Sledgehammer}~\cite{sledgehammer}, 
which searches for proofs by calling external automated theorem provers, and then converts the proofs discovered to Isar. Two counterexample finding tools, \textit{nitpick} and \textit{Quickcheck}, are also provided. Moreover, the command \isa{try0} calls a number of built-in proof methods (such as simplification) in search of a proof. 
 The user interface is Isabelle/jEdit, which provides real-time proof checking, as well as rich semantic information for the formal text and direct links to the user manuals and tutorials.\footnote{These can also be found under ``Documentation'' on the Isabelle webpage.}
We have written an introduction to Isabelle aimed at mathematicians \cite{praxis}.

The Isabelle/HOL libraries and 
the AFP contain a vast amount of formalized material, covering many areas of pure mathematics as well as in computer science, logic and even philosophy.

\section{On the irrationality of certain series}
\subsection{Results formalized}
The following results are by Erd\H{o}s and Straus \cite{erdosstraus}.
\begin{theorem}[\hbox{Erd\H{o}s and Straus \cite[Theorem 2.1]{erdosstraus}}]
\label{thm:erdos_straus_thm_2.1}
Let $\{ b_n\}^\infty_{n=1}$ be a sequence of integers and
$\{ \alpha_n\}^\infty_{n=1}$ a sequence of positive integers with $\alpha_n >1$ for all large $n$ and 
$$\lim_{n=1, n \rightarrow \infty} \frac{|b_n|}{\alpha_{n-1}\alpha_n} =0.$$ Then the sum $$\sum^\infty_{n=1} \frac{b_n}
{ \prod^n_{i=1} \alpha_i}$$ is rational if and only if there exists a positive integer $B$ and 
 a sequence of integers $\{ c_n\}^\infty_{n=1}$ so that for all large $n$,
\begin{equation*} \label{eq:erdos_straus_Bc}
    B b_n =c_n \alpha_n -c_{n+1},~~ |c_{n+1}| <\alpha_n/2. 
\end{equation*} 

\end{theorem}

\begin{corollary}[\hbox{Erd\H{o}s and Straus \cite[Corollary 2.10]{erdosstraus}}] Let $\{ \alpha_n\}^\infty_{n=1}$ and $\{ b_n\}^\infty_{n=1}$ satisfy the hypotheses of the theorem above and in addition that for all large $n$ we have $b_n>0$, $\alpha_{n+1} \geq \alpha_n$, $\lim_{n \rightarrow \infty} (b_{n+1} - b_n)/\alpha_n \leq 0$ and $\liminf_{n \rightarrow \infty}  \alpha_n /b_n =0$. Then the sum $$\sum^\infty_{n=1} \frac{b_n}{\prod^n_{i=1} \alpha_i }$$ is irrational.
\end{corollary}
\begin{theorem}[\hbox{Erd\H{o}s and Straus \cite[Theorem 3.1] {erdosstraus}}]
\label{thm:erdos_straus_thm_3.1}

Let $p_n$ be the $n$th prime number and let 
$\{ \alpha_n\}^\infty_{n=1}$ be a monotonic sequence of positive integers satisfying 
$\lim_{n \rightarrow \infty} p_n/\alpha_n^2 =0$ and $\liminf_{n \rightarrow \infty} \alpha_n/p_n =0$.  Then the sum $$\sum^\infty_{n=1} \frac{p_n}{\prod^n_{i=1} \alpha_i} $$ is irrational.
\end{theorem}

\subsection{On the formalization}
Our full formalization is online \cite{irrationalityerdosstraus} and here we discuss certain key points. The formalization process turned out to be time-consuming: we had to fill in a number of intermediate reasoning steps that had to be made explicit in the formal proofs.
For example, in the proof of Theorem 3.2 (\cite[Theorem 3.1]{erdosstraus}), for the sequence of integers
$\{c_n\}^\infty_{n=1}$ (for which $c_n>0$ for large $n$), 
Erd\H{o}s and Straus claim that since it is unbounded, for large $n$ there must exist an index $m \geq n$ so that 
$c_m \leq c_n < c_{m+1}$. In order to prove this claim, we had first to show that 
for large $n$ the sequence $\{c_n\}^\infty_{n=1}$ is neither monotone increasing nor monotone decreasing, which required some work.

We moreover had to do a certain amount of restructuring for the formalization of the proof of Theorem 3.1 (\cite[Theorem 2.1]{erdosstraus}), as the original proof is based on a sketch of a construction for obtaining the terms of the desired sequence 
$\{c_n\}^\infty_{n=1}$, starting with the construction of the first two terms for large enough $n$ (``Proceeding in this manner we get the desired sequence''). In particular, in the sufficient direction, we need to obtain a positive integer $B$ and a sequence of integers $\{c_n\}^\infty_{n=1}$ that satisfy $B b_n =c_n \alpha_n -c_{n+1}$ and $|c_{n+1}| <\alpha_n/2$ for all large enough $n$, where $\{b_n\}^\infty_{n=1}$ and $\{\alpha_n\}^\infty_{n=1}$ are sequences of integers as in the assumptions of the theorem. In the middle of the proof, we will have for a large enough $N$
\begin{equation} \label{eq:RN_integer}
    \frac{B b_N}{\alpha_N} - R_N \textrm{ is an integer,}
\end{equation}
where $\{R_n\}^\infty_{n=1}$ is a sequence such that for all $n \geq N$, $| R_n | < 1/4$ and  
\begin{equation} \label{eq:R_inductive}
    R_{n+1} = \alpha_n R_n - \frac{B b_{n+1}}{\alpha_{n+1}}.
\end{equation}
To construct a suitable $\{c_n\}^\infty_{n=1}$, we let $c_N$ be the integer nearest to $(B b_N)/\alpha_N$ and inductively construct $c_{n+1} = c_n \alpha_n - B b_n$ (for all $n \geq N$). This was achieved by defining a recursive function in Isabelle/HOL:
\begin{isabelle}
\ \isacommand{fun}\isamarkupfalse%
\ get{\isacharunderscore}c{\isacharcolon}{\isacharcolon}{\isachardoublequoteopen}{\isacharparenleft}nat\ {\isasymRightarrow}\ int{\isacharparenright}\ {\isasymRightarrow}\ {\isacharparenleft}nat\ {\isasymRightarrow}\ int{\isacharparenright}\ \isanewline
\ \ \ \ \ \ \ \ \ \ \ \ \ \ \ \ \ \ \ \ \ \ \ \ \ \ {\isasymRightarrow}\ int\ {\isasymRightarrow}\ nat\ {\isasymRightarrow}\ {\isacharparenleft}nat\ {\isasymRightarrow}\ int{\isacharparenright}{\isachardoublequoteclose}\ \isakeyword{where}\isanewline
\ \ {\isachardoublequoteopen}get{\isacharunderscore}c\ a{\isacharprime}\ b{\isacharprime}\ B\ N\ {\isadigit{0}}\ {\isacharequal}\ round\ {\isacharparenleft}B\ {\isacharasterisk}\ b{\isacharprime}\ N\ {\isacharslash}\ a{\isacharprime}\ N{\isacharparenright}{\isachardoublequoteclose}{\isacharbar}\isanewline
\ \ {\isachardoublequoteopen}get{\isacharunderscore}c\ a{\isacharprime}\ b{\isacharprime}\ B\ N\ {\isacharparenleft}Suc\ n{\isacharparenright}\ {\isacharequal}\ get{\isacharunderscore}c\ a{\isacharprime}\ b{\isacharprime}\ B\ N\ n\ {\isacharasterisk}\ a{\isacharprime}\ {\isacharparenleft}n{\isacharplus}N{\isacharparenright}\ \isanewline
\ \ \ \ \ \ \ \ \ \ \ \ \ \ \ \ \ \ \ \ \ \ \ \ \ \ \ \ \ \ \ \ \ \ \ \ \ \ \ \ \ \ \ \ \ \ \ {\isacharminus}\ B\ {\isacharasterisk}\ b{\isacharprime}\ {\isacharparenleft}n{\isacharplus}N{\isacharparenright}{\isachardoublequoteclose}
\end{isabelle}
where \isa{round} returns the nearest integer of an input number. By construction, we already have $B b_n =c_n \alpha_n -c_{n+1}$ so the goal is to show $|c_{n+1}| <\alpha_n/2$ (for all $n \geq N$). To achieve this, we deploy proof by induction to derive 
\[
    c_n - (B b_n) / \alpha_n = R_n,
\]
where (\ref{eq:RN_integer}) has been used for the base case and (\ref{eq:R_inductive}) for the inductive one. We can then close the proof by having 
\[
    \frac{|c_{n+1}|}{\alpha_n} = |R_n| < \frac{1}{2}.
\]
The original proof repeatedly invokes variants of (\ref{eq:RN_integer}) and tries to prove $c_n$ is the closest integer to $(B b_n) / \alpha_n$ for all $n \geq N$ (e.g., (2.7)-(2.9) in the original proof). We believe our altered proof might be more straightforward.

We also noted that in the original proofs of
Corollary 3.1 (\cite[Corollary 2.10]{erdosstraus}) and Theorem 3.2 (\cite[Theorem 3.1]{erdosstraus}) 
certain inequalities required some minor corrections (for details we refer to our formalization \cite{irrationalityerdosstraus}), which fortunately did not affect the correctness of the statements nor the original proof structures.

Our formalized versions of the statements of  
Theorem 3.1 (\cite[Theorem 2.1]{erdosstraus}),
Corollary 3.1 (\cite[Corollary 2.10]{erdosstraus}) and Theorem 3.2 (\cite[Theorem 3.1]{erdosstraus}) read as follows:
\begin{isabelle}
\isacommand{theorem}\isamarkupfalse%
\ theorem{\isacharunderscore}{\isadigit{2}}{\isacharunderscore}{\isadigit{1}}{\isacharunderscore}Erdos{\isacharunderscore}Straus{\isacharcolon}\isanewline
\ \ \isakeyword{fixes}\ a\ b\ {\isacharcolon}{\isacharcolon}\ {\isachardoublequoteopen}nat{\isasymRightarrow}int{\isachardoublequoteclose}\isanewline
\ \ \isakeyword{assumes}\ {\isachardoublequoteopen}{\isasymforall}\ n{\isachardot}\ a\ n\ {\isachargreater}{\isadigit{0}}{\isachardoublequoteclose}\ \isakeyword{and}\ {\isachardoublequoteopen}{\isasymforall}\isactrlsub F\ n\ in\ sequentially{\isachardot}\ a\ n\ {\isachargreater}\ {\isadigit{1}}{\isachardoublequoteclose}\ \isanewline
\ \ \ \ \isakeyword{and}\ {\isachardoublequoteopen}{\isacharparenleft}{\isasymlambda}n{\isachardot}\ {\isasymbar}b\ n{\isasymbar}\ {\isacharslash}\ {\isacharparenleft}a\ {\isacharparenleft}n{\isacharminus}{\isadigit{1}}{\isacharparenright}{\isacharasterisk}a\ n{\isacharparenright}{\isacharparenright}\ {\isasymlonglonglongrightarrow}\ {\isadigit{0}}{\isachardoublequoteclose}\isanewline
\ \ \isakeyword{shows}\ {\isachardoublequoteopen}{\isacharparenleft}{\isasymSum}n{\isachardot}\ {\isacharparenleft}b\ n\ {\isacharslash}\ {\isacharparenleft}{\isasymProd}i\ {\isasymle}\ n{\isachardot}\ a\ i{\isacharparenright}{\isacharparenright}{\isacharparenright}\ {\isasymin}\ {\isasymrat}\ {\isasymlongleftrightarrow}\ \isanewline
\ \ \ \ \ \ \ \ \ \ \ \ \ \ {\isacharparenleft}{\isasymexists}\ {\isacharparenleft}B{\isacharcolon}{\isacharcolon}int{\isacharparenright}{\isachargreater}{\isadigit{0}}{\isachardot}\ {\isasymexists}\ c{\isacharcolon}{\isacharcolon}nat{\isasymRightarrow}\ int{\isachardot}\isanewline
\ \ \ \ \ \ \ \ \ \ \ \ \ \ \ \ \ \ \ \ \ {\isacharparenleft}{\isasymforall}\isactrlsub F\ n\ in\ sequentially{\isachardot}\ B{\isacharasterisk}b\ n\ {\isacharequal}\ c\ n\ {\isacharasterisk}\ a\ n\ {\isacharminus}\ c{\isacharparenleft}n{\isacharplus}{\isadigit{1}}{\isacharparenright}\isanewline
\ \ \ \ \ \ \ \ \ \ \ \ \ \ \ \ \ \ \ \ \ \ \ \ \ \ \ \ \ \ \ \ \ \ \ \ \ \ \ \ \ \ \ \ \ \ \ \ \ \ \ \ {\isasymand}\ {\isasymbar}c{\isacharparenleft}n{\isacharplus}{\isadigit{1}}{\isacharparenright}{\isasymbar}{\isacharless}a\ n{\isacharslash}{\isadigit{2}}{\isacharparenright} {\isacharparenright}{\isachardoublequoteclose}
\end{isabelle}

\begin{isabelle}
\isacommand{corollary}\isamarkupfalse%
\ corollary{\isacharunderscore}{\isadigit{2}}{\isacharunderscore}{\isadigit{1}}{\isadigit{0}}{\isacharunderscore}Erdos{\isacharunderscore}Straus{\isacharcolon}\isanewline
\ \ \isakeyword{fixes}\ a\ b\ {\isacharcolon}{\isacharcolon}\ {\isachardoublequoteopen}nat{\isasymRightarrow}int{\isachardoublequoteclose}\isanewline
\ \ \isakeyword{assumes}\ {\isachardoublequoteopen}{\isasymforall}\ n{\isachardot}\ a\ n\ {\isachargreater}{\isadigit{0}}{\isachardoublequoteclose}\ \isakeyword{and}\ {\isachardoublequoteopen}{\isasymforall}\isactrlsub F\ n\ in\ sequentially{\isachardot}\ a\ n\ {\isachargreater}\ {\isadigit{1}}{\isachardoublequoteclose}\ \isanewline
\ \ \ \ \isakeyword{and}\ {\isachardoublequoteopen}{\isacharparenleft}{\isasymlambda}n{\isachardot}\ {\isasymbar}b\ n{\isasymbar}\ {\isacharslash}\ {\isacharparenleft}a\ {\isacharparenleft}n{\isacharminus}{\isadigit{1}}{\isacharparenright}{\isacharasterisk}a\ n{\isacharparenright}{\isacharparenright}\ {\isasymlonglonglongrightarrow}\ {\isadigit{0}}{\isachardoublequoteclose}\isanewline
\ \ \ \ \isakeyword{and}\ {\isachardoublequoteopen}{\isasymforall}\isactrlsub F\ n\ in\ sequentially{\isachardot}\ b\ n\ {\isachargreater}\ {\isadigit{0}}\ {\isasymand}\ a\ {\isacharparenleft}n{\isacharplus}{\isadigit{1}}{\isacharparenright}\ {\isasymge}\ a\ n{\isachardoublequoteclose}\ \isanewline
\ \ \ \ \isakeyword{and}\ {\isachardoublequoteopen}lim\ {\isacharparenleft}{\isasymlambda}n{\isachardot}\ {\isacharparenleft}b{\isacharparenleft}n{\isacharplus}{\isadigit{1}}{\isacharparenright}\ {\isacharminus}\ b\ n{\isacharparenright}\ {\isacharslash}\ a\ n{\isacharparenright}\ {\isasymle}\ {\isadigit{0}}{\isachardoublequoteclose}\ \isanewline
\ \ \ \ \isakeyword{and}\ {\isachardoublequoteopen}convergent\ {\isacharparenleft}{\isasymlambda}n{\isachardot}\ {\isacharparenleft}b{\isacharparenleft}n{\isacharplus}{\isadigit{1}}{\isacharparenright}\ {\isacharminus}\ b\ n{\isacharparenright}\ {\isacharslash}\ a\ n{\isacharparenright}{\isachardoublequoteclose}\isanewline
\ \ \ \ \isakeyword{and}\ {\isachardoublequoteopen}liminf\ {\isacharparenleft}{\isasymlambda}n{\isachardot}\ a\ n\ {\isacharslash}\ b\ n{\isacharparenright}\ {\isacharequal}\ {\isadigit{0}}\ {\isachardoublequoteclose}\isanewline
\ \ \isakeyword{shows}\ {\isachardoublequoteopen}{\isacharparenleft}{\isasymSum}n{\isachardot}\ {\isacharparenleft}b\ n\ {\isacharslash}\ {\isacharparenleft}{\isasymProd}i\ {\isasymle}\ n{\isachardot}\ a\ i{\isacharparenright}{\isacharparenright}{\isacharparenright}\ {\isasymnotin}\ {\isasymrat}{\isachardoublequoteclose}
\end{isabelle}

\begin{isabelle}
\isacommand{theorem}\ theorem\_3\_10\_Erdos\_Straus:\isanewline
\ \ \isakeyword{fixes}\ a::"nat\ \isasymRightarrow \ int"\isanewline
\ \ \isakeyword{assumes}\ "\isasymforall \ n.\ a\ n\ >0"\ \isakeyword{and}\ "mono\ a"\isanewline
\ \ \ \ \isakeyword{and}\ "(\isasymlambda n.\ nth\_prime\ n\ /\ (a\ n)\isacharcircum 2)\ \isasymlonglonglongrightarrow \ 0"\isanewline
\ \ \ \ \isakeyword{and}\ "liminf\ (\isasymlambda n.\ a\ n\ /\ nth\_prime\ n)\ =\ 0"\isanewline
\ \ \isakeyword{shows}\ "(\isasymSum n.\ (nth\_prime\ n\ /\ (\isasymProd i\ \isasymle \ n.\ a\ i)))\ \isasymnotin \ \isasymrat "
\end{isabelle}
Above, \isa{sequentially} is a \emph{filter} \cite{isa_filter} for expressing limits indexed by positive integers. 
So \isa{\isasymforall \isactrlsub F\ n\ in\ sequentially.\ P\ n} expresses that \isa{P n} holds for all sufficiently large~\isa{n}.

\subsubsection{Calculations with real asymptotics} 
Asymptotic arguments (e.g.\ limits and statements
of the form ``for all sufficiently large~$n$'') present real difficulties for mechanised proofs.
Justifications omitted as obvious in paper proofs need to be written out. For example, in the proof of Theorem 3.2 (\cite[Theorem 3.1]{erdosstraus})
 we need to formally prove
\begin{equation} \label{eq:asym_ex}
    8 B \ln{n} + 1 < \frac{\sqrt{n}}{4} \qquad \text{for all large $n$},
\end{equation}
where $B$ is a positive constant. Since the proof of Eq.\ (\ref{eq:asym_ex}) involves routine calculations of limits, this step is completely omitted in the paper proof. In Isabelle/HOL, Eq.\ (\ref{eq:asym_ex}) is encoded as
\begin{isabelle}
"\isasymforall \isactrlsub F\ n\ in\ sequentially.\ 8*B*ln\ n\ +\ 1<sqrt\ n/4"
\end{isabelle}
Our first attempt resulted in the following proof:
\begin{isabelle}
\ \ \ \ \ \ \isacommand{have}\ "\isasymforall \isactrlsub F\ n\ in\ sequentially.\ 8*B*ln\ n\ +\ 1<sqrt\ n/4"\isanewline
\ \ \ \ \ \ \isacommand{proof}\ -\isanewline
\ \ \ \ \ \ \ \ \isacommand{have}\ "(\isasymlambda n.\ ln\ n\ /\ sqrt\ n)\ \isasymlonglonglongrightarrow \ 0"\isanewline
\ \ \ \ \ \ \ \ \ \ \isacommand{using}\ lim\_ln\_over\_power[of\ "1/2"]\ powr\_half\_sqrt\ \isacommand{by}\ simp\isanewline
\ \ \ \ \ \ \ \ \isacommand{from}\ tendsto\_mult\_right\_zero[OF\ this]\ \isanewline
\ \ \ \ \ \ \ \ \isacommand{have}\ "(\isasymlambda n.\ 8\ * B\ *\ \ ln\ n\ /\ sqrt\ n)\ \isasymlonglonglongrightarrow \ 0"\ \isacommand{by}\ auto\isanewline
\ \ \ \ \ \ \ \ \isacommand{moreover}\ \isacommand{have}\ "(\isasymlambda n.\ 1\ /\ sqrt\ n)\ \isasymlonglonglongrightarrow \ 0"\isanewline
\ \ \ \ \ \ \ \ \ \ \isacommand{apply}\ (rule\ real\_tendsto\_divide\_at\_top)\isanewline
\ \ \ \ \ \ \ \ \ \ \isacommand{by}\ (auto\ simp\ add:\ filterlim\_of\_nat\_at\_top\ sqrt\_at\_top)\isanewline
\ \ \ \ \ \ \ \ \isacommand{ultimately}\ \isacommand{have}\ "(\isasymlambda n.\ 8*B*ln\ n/sqrt\ n\ +\ 1/sqrt\ n)\ \isasymlonglonglongrightarrow \ 0"\isanewline
\ \ \ \ \ \ \ \ \ \ \isacommand{by}\ (auto\ intro:tendsto\_add\_zero)\isanewline
\ \ \ \ \ \ \ \ \isacommand{from}\ tendstoD[OF\ this,\ of\ "1/4"]\isanewline
\ \ \ \ \ \ \ \ \isacommand{have}\ "\isasymforall \isactrlsub F\ n\ in\ sequentially.\ 8*B*ln\ n/sqrt\ n\ +\ 1/sqrt\ n\ <\ 1/4"\isanewline
\ \ \ \ \ \ \ \ \ \ \isacommand{unfolding}\ abs\_less\_iff\ dist\_real\_def\ \isanewline
\ \ \ \ \ \ \ \ \ \ \isacommand{by}\ (auto\ elim!:eventually\_mono)\isanewline
\ \ \ \ \ \ \ \ \isacommand{then}\ \isacommand{show}\ ?thesis\isanewline
\ \ \ \ \ \ \ \ \ \ \isacommand{using}\ eventually\_gt\_at\_top[of\ 0]\ \isanewline
\ \ \ \ \ \ \ \ \ \ \isacommand{by}\ (eventually\_elim,\ auto\ simp:field\_simps)\isanewline
\ \ \ \ \ \ \isacommand{qed}
\end{isabelle}
The proof is tedious. Let's work through the text above. We start with primitive properties of the ln and square root functions, 
\[
    \lim_{n \rightarrow \infty} \frac{\ln{n}}{\sqrt{n}} = 0\qquad \mathrm{and} \qquad \lim_{n \rightarrow \infty}  \frac{1}{\sqrt{n}} = 0,
\]
written \isa{"(\isasymlambda n.\ ln\ n\ /\ sqrt\ n)\ \isasymlonglonglongrightarrow \ 0"} and \isa{"(\isasymlambda n.\ 1\ /\ sqrt\ n)\ \isasymlonglonglongrightarrow \ 0"} in the script. Then we need to manually transform a property of limits,
\[
    \lim_{n \rightarrow \infty} \frac{8 B \ln{n}}{ \sqrt{n}} + \frac{1}{\sqrt{n}} = 0
\]
to the proposition that
\[
    \frac{8 B \ln{n}}{ \sqrt{n}} + \frac{1}{\sqrt{n}} < \frac{1}{4} \qquad \textrm{for all large enough $n$},
\]
by applying the lemma \isa{tendstoD}:
\[
\text{if }  \lim_{x \rightarrow F} f(x) = l 
\text{ and } 0 < e \text{ then eventually\footnotemark}
   \left| f(x) - l \right| < e.
\]
\footnotetext{as $x$ approaches the filter $F$}%
Fortunately, Manuel Eberl \cite{asym_eberl} had just implemented the tactic \isa{real\_asymp} that solves Eq.\ (\ref{eq:asym_ex}) automatically, so we can reduce our initial proof to just one line: 
\begin{isabelle}
\ \ \ \ \ \ \isacommand{have}\isamarkupfalse%
\ {\isachardoublequoteopen}{\isasymforall}\isactrlsub F\ n\ in\ sequentially{\isachardot}\ {\isadigit{8}}{\isacharasterisk}B{\isacharasterisk}ln\ n\ {\isacharplus}\ {\isadigit{1}}{\isacharless}sqrt\ n{\isacharslash}{\isadigit{4}}{\isachardoublequoteclose}\isanewline
\ \ \ \ \ \ \ \ \isacommand{by} real\_asymp
\end{isabelle}
The basic idea behind \isa{real\_asymp} is to approximate a continuous function $g: \mathbb{R} \rightarrow \mathbb{R}$ when its argument approaches a limit $L$ using Poincar\'{e} expansion (or asymptotic series):
\[
    \sum_{n=1}^N a_n \varphi_n (x),
\]
where $a_n \in \mathbb{R}$ and $\{\varphi_n(x)\}$ is a sequence of continuous functions such that $\varphi_{n+1}(x) \in o(\varphi_n(x))$ for all $n$ as $x \rightarrow L$.\footnote{Here, $o(\cdot)$ is the `small o' notation, and $\varphi_{n+1}(x) \in o(\varphi_n(x))$ as $x \rightarrow L$ is equivalent to $\lim_{x \rightarrow L} \varphi_{n+1}(x) / \varphi_n(x) = 0$. }
Through arithmetic on the formal series, the asymptotic behaviour of a function at $L$ can then be decided. In general, \isa{real\_asymp} handles functions built from basic arithmetic and elementary functions including \isa{exp}, \isa{ln}, and \isa{sin}.

In situations involving limits, due to the presence of bound variables,
Sledgehammer's automation helps little, so \isa{real\_asymp} makes a real contribution.
Such domain-specific automation inspired from computer algebra systems will surely prove to be of great use when mechanising mathematical proofs.

But in general we have found Sledgehammer to be of tremendous value. A telltale sign that it has been used is any occurrence of the proof methods \isa{metis} or \isa{smt}, which are littered throughout our formal proofs \cite{irrationalityhancl, transhanclrucki, irrationalityerdosstraus}. These methods are difficult to use manually, and any instances of them are almost certain to have been generated by Sledgehammer.

\subsubsection{Reasoning with prime numbers} 

Elementary properties concerning prime numbers are often assumed in informal proofs, with supporting calculations largely omitted. Thanks to the mechanised proof of the prime number theorem (PNT) \cite{PrimeNumberTheorem} as well as another development on primes \cite{primefacts} from the AFP, basic reasoning with prime numbers is feasible. However, there are still some tedious, repetitious calculations, which could be optimised in the future. For example, the first step in proving Theorem \ref{thm:erdos_straus_thm_3.1} (\cite[Theorem 3.1]{erdosstraus}) is to demonstrate that the hypotheses of Theorem \ref{thm:erdos_straus_thm_2.1} (\cite[Theorem 2.1]{erdosstraus}) are satisfied by the former's assumptions.
Although this step has been completely omitted in the informal proof, we need to explicitly derive 
$\lim_{n \rightarrow \infty} {p_n}/(\alpha_{n-1} \alpha_n)$
from $\lim_{n \rightarrow \infty} {p_n}/{\alpha_n^2}$,
where $p_n$ is the $n^{\text{th}}$ prime number and $\{\alpha_n\}^\infty_{n=1}$ is a monotonic sequence of positive integers. The derivation works as follows:
\[
\lim_{n \rightarrow \infty} \frac{p_n}{\alpha_{n-1} a_n} \leq \lim_{n \rightarrow \infty} \frac{p_n}{\alpha_{n-1}^2} \leq  \lim_{n \rightarrow \infty} \frac{2 p_{n-1}}{\alpha_{n-1}^2} 
= 2\lim_{n \rightarrow \infty} \frac{p_n}{\alpha_n^2} = 0.
\]

The second inequality requires a basic property of the prime numbers:
\begin{equation} \label{eq:consecutive_primes}
    \lim_{n \rightarrow \infty} \frac{p_{n+1}}{p_n} = 1.
\end{equation}
Starting with a form of the PNT, namely $p_n \sim n \ln(n) $, this property alone takes 28 lines of tedious limit calculations similar to those in our first attempt to prove (\ref{eq:asym_ex}). Another example 
from the proof of Theorem \ref{thm:erdos_straus_thm_3.1} (\cite[Theorem 3.1]{erdosstraus})  
is the following: considering the interval $ [N,2N)$ for large enough $N$ and given that $c_{n+1} > c_n$ implies $p_{n+1} > p_n + \sqrt{p_n}$, we need to deduce that $c_{n+1} > c_n$ happens for fewer than half of the integers within $[N,2N)$. The derivation works because
\begin{equation} \label{eq:nth_prime_double_sqrt_less}
    \frac{p_{2 N} - p_N}{\sqrt{p_N}} < N^{\frac{1}{2} + \epsilon} \qquad \textrm{for all $\epsilon>0$ and all large enough $N$,}
\end{equation}
and $N^{\frac{1}{2} + \epsilon} < N / 2$ when $\epsilon < 1/2$ and $N$ is large enough. Property (\ref{eq:nth_prime_double_sqrt_less}) is formulated in Isabelle/HOL as 
\begin{isabelle}
\isacommand{lemma} \ nth\_prime\_double\_sqrt\_less:\isanewline
\ \ \isakeyword{assumes}\ "\isasymepsilon \ >\ 0"\isanewline
\ \ \isakeyword{shows}\ "\isasymforall \isactrlsub F\ n\ in\ sequentially.\ (nth\_prime\ (2*n)\ -\ nth\_prime\ n)\ \isanewline
\ \ \ \ \ \ \ \ \ \ \ \ /\ sqrt\ (nth\_prime\ n)\ <\ n\ powr\ (1/2\ +\ \isasymepsilon )"
\end{isabelle}
where we remind that \isa{{\isasymforall}\isactrlsub F\ n\ in\ sequentially} stands for the ``large enough'' quantifier. Despite (\ref{eq:nth_prime_double_sqrt_less}) being merely claimed without proof in the paper, it took us 56 lines to prove it in Isabelle/HOL starting from the PNT\@. In short, we are happy that properties related to the prime numbers are now viable in Isabelle/HOL, but we wish we could derive their asymptotic properties like (\ref{eq:consecutive_primes}) and~(\ref{eq:nth_prime_double_sqrt_less}) with fewer tears. 
Automating this process through the tactic \isa{real\_asymp} is desirable but the tactic does not yet support the function \isa{nth\_prime}: in contrast to elementary functions like \isa{exp}, \isa{nth\_prime} cannot be asymptotically expanded to an arbitrary order. Incorporating \isa{nth\_prime} into \isa{real\_asymp} may require special treatment.

\section{Irrational rapidly convergent series}

\subsection{Results formalized}
The following results are by Han\v{c}l \cite{hancl}.
\begin{theorem}[\hbox{Han\v{c}l \cite[Theorem 3]{hancl}}] Let $A>1$ be a real number. Let $\{d_n\}^\infty_{n=1}$ be a sequence of real numbers greater than one. Let $\{ \alpha_n\}^\infty_{n=1}$ and $\{ b_n\}^\infty_{n=1}$ be sequences of positive integers such that $\lim_{n \rightarrow \infty} \alpha_n^{\frac{1}{2^n}} = A$
and for all sufficiently large $n$ 
$$\frac{A}{ \alpha_n^{\frac{1}{2^n}}} > \prod^\infty_{j=n} d_j
\quad\text{and}\quad \lim_{n \rightarrow \infty}  \frac{d_n^{2^n}}{b_n} = \infty.$$
Then the sum $$\sum^\infty_{n=1} \frac{b_n}{\alpha_n}$$ is irrational.
\end{theorem}

\begin{corollary}[\hbox{Han\v{c}l \cite[Corollary 2]{hancl}}]Let $A>1$ and let $\{ \alpha_n\}^\infty_{n=1}$ and $\{ b_n\}^\infty_{n=1}$ be sequences of positive integers such that $\lim_{n \rightarrow \infty} \alpha_n^{\frac{1}{2^n}} =A$. Then, assuming that for every sufficiently large positive integer $n$, $ \alpha_n^{\frac{1}{2^n}} (1+ 4(2/3)^n)\leq A$ and $b_n \leq 2^{(4/3)^{n-1}}$, the sum $\sum^\infty_{n=1} b_n /\alpha_n$ is irrational.
\end{corollary}
This corollary follows directly from the theorem by an appropriate choice of the sequence $\{d_n\}^\infty_{n=1}$, in particular by choosing  
$d_n = 1+(2/3)^n$ so that the theorem can be directly applied.

\subsection{On the formalization}

Our full formalization is online on the AFP \cite{irrationalityhancl}. 
This was the first work in this area that we formalized, starting in early 2018. We chose
this research paper as the proofs are interesting but not too complicated, and even though this is a topic 
in number theory, they are based on fundamental material from analysis. 
Of course, we also had to show quite a few auxiliary facts, including lemmas on summability criteria for series. For instance, we modified the classic ratio test for series (which was already available in the HOL Library):
\begin{isabelle}
\isacommand{lemma}\ summable\_ratio\_test:\isanewline
\ \ \isakeyword{fixes}\ c\ ::\ real\ \isakeyword{and}\ N\ ::\ nat\ \isanewline
\ \ \ \ \isakeyword{and}\ f\ ::\ "nat\ \isasymRightarrow \ 'a\ ::\ real\_normed\_vector"\isanewline
\ \ \isakeyword{assumes}\ "c\ <\ 1"\ \isakeyword{and}\ "\isasymforall n\ \isasymge \ N.\ norm\ (f\ (Suc\ n))\ \isasymle \ c\ *\ norm\ (f\ n)"\isanewline
\ \ \isakeyword{shows}\ "summable\ f"
\end{isabelle}
The version modified in terms of limits is given below:
\begin{isabelle}
\isacommand{lemma}\ summable\_ratio\_test\_tendsto:\isanewline
\ \ \isakeyword{fixes}\ c\ ::\ real\ \isakeyword{and}\ f\ ::\ "nat\ \isasymRightarrow \ 'a\ ::\ real\_normed\_vector"\isanewline
\ \ \isakeyword{assumes}\ "c\ <\ 1"\ \isakeyword{and}\ "\isasymforall n.\ f\ n\ \isasymnoteq \ 0"\ \isanewline
\ \ \ \ \isakeyword{and}\ tendsto\_c:\ "(\isasymlambda n.\ norm\ (f\ (Suc\ n))\ /\ norm\ (f\ n))\ \isasymlonglonglongrightarrow \ c"\isanewline
\ \ \isakeyword{shows}\ "summable\ f"\isanewline
\isacommand{proof}\ -\isanewline
\ \ \isacommand{from}\ \isacartoucheopen c<1\isacartoucheclose \ tendsto\_c\isanewline
\ \ \isacommand{obtain}\ N\ \isakeyword{where}\ N\_dist:\isanewline
\ \ \ \ \ \ "\isasymforall n\isasymge N.\ dist\ (norm\ (f\ (Suc\ n))\ /\ norm\ (f\ n))\ c\ <\ (1-c)/2"\isanewline
\ \ \ \ \isacommand{unfolding}\ tendsto\_iff\ eventually\_sequentially\ \isanewline
\ \ \ \ \isacommand{by}\ (meson\ diff\_gt\_0\_iff\_gt\ zero\_less\_divide\_iff\ zero\_less\_numeral)\isanewline
\ \ \isacommand{have}\ "norm\ (f\ (Suc\ n))\ /\ norm\ (f\ n)\ \isasymle \ (1+c)/2"\ \isakeyword{when}\ "n\isasymge N"\ \isakeyword{for}\ n\isanewline
\ \ \ \ \isacommand{using}\ N\_dist[rule\_format,OF\ \isacartoucheopen n\isasymge N\isacartoucheclose ]\ \isacartoucheopen c<1\isacartoucheclose \ \isanewline
\ \ \ \ \isacommand{by}\ (auto\ simp\ add:field\_simps\ dist\_norm,argo)\isanewline
\ \ \isacommand{then}\ \isacommand{have}\ "norm\ (f\ (Suc\ n))\ \isasymle \ (1+c)/2\ *\ norm\ (f\ n)"\ \isanewline
\ \ \ \ \ \ \isakeyword{when}\ "n\isasymge N"\ \isakeyword{for}\ n\isanewline
\ \ \ \ \isacommand{using}\ \isacartoucheopen n\isasymge N\isacartoucheclose \ \isacartoucheopen \isasymforall n.\ f\ n\ \isasymnoteq \ 0\isacartoucheclose \ \isacommand{by}\ (auto\ simp\ add:divide\_simps)\isanewline
\ \ \isacommand{moreover}\ \isacommand{have}\ "(1+c)/2\ <\ 1"\ \isacommand{using}\ \isacartoucheopen c<1\isacartoucheclose \ \isacommand{by}\ auto\ \isanewline
\ \ \isacommand{ultimately}\ \isacommand{show}\ "summable\ f"\isanewline
\ \ \ \ \isacommand{using}\ summable\_ratio\_test\ \isacommand{by}\ blast\isanewline
\isacommand{qed}
\end{isabelle}

Our formalized version of Theorem 4.1 (\cite[Theorem 3]{hancl}) is presented below. We include its formalized proof in the Appendix.
The reader will notice that the formal proof is only around 50 lines of Isar code;
however, it comes after about 800 lines of auxiliary lemmas that are required for this specific proof. 
As in the original---and after having established the summability of the series in question---the proof is by contradiction.
Assuming that the sum is rational, for a quantity 
 $\textrm{ALPHA}(n)$ we show that $\textrm{ALPHA}(n) \geq 1$ for all $n \in \mathbb{N}$. After that, we show the existence of an $n \in \mathbb{N}$ for which $\textrm{ALPHA}(n) < 1$, a contradiction: hence, the sum of the series is irrational. 
 As already mentioned, Isar admits structured proofs, which 
are more readable than the usual sequence of proof commands. In our proof, the structure of a proof by contradiction is visible:

\isa{
\isacommand{proof}\isamarkupfalse%
{\isacharparenleft}rule\ ccontr{\isacharparenright} [...] \isacommand{show} False\ [...]
\isacommand{qed}}.

\begin{isabelle}
\isacommand{theorem}\ Hancl3:\isanewline
\ \ \isakeyword{fixes}\ d\ ::"nat\isasymRightarrow real"\ \isakeyword{and}\ \ a\ b\ ::\ "nat\isasymRightarrow int"\isanewline
\ \ \isakeyword{assumes}\ "A\ >\ 1"\ \isakeyword{and}\ d:\ "\isasymforall n.\ d\ n\ >\ 1"\isanewline
\ \ \ \ \isakeyword{and}\ a:\ "\isasymforall n.\ a\ n>0"\ \isakeyword{and}\ b:\ "\isasymforall n.\ b\ n\ >\ 0"\ \isakeyword{and}\ "s>0"\isanewline
\ \ \ \ \isakeyword{and}\ "(\isasymlambda n.\ (a\ n)\ powr(1\ /\ of\_int(2\isacharcircum n)))\ \isasymlonglonglongrightarrow \ A"\isanewline
\ \ \ \ \isakeyword{and}\ "\isasymforall n\ \isasymge \ s.\ A\ /\ (a\ n)\ powr\ (1\ /\ of\_int(2\isacharcircum n))\ >\ (\isasymProd j.\ d\ (n+j))"\isanewline
\ \ \ \ \isakeyword{and}\ "LIM\ n\ sequentially.\ d\ n\ \isacharcircum \ 2\ \isacharcircum \ n\ /\ b\ n\ :>\ at\_top"\isanewline
\ \ \ \ \isakeyword{and}\ "convergent\_prod\ d"\ \isanewline
\ \ \isakeyword{shows}\ "(\isasymSum n.\ b\ n\ /\ a\ n)\ \isasymnotin \ \isasymrat 
\end{isabelle}

Our formalized version of Corollary 4.1 (\cite[Corollary 2]{hancl}) is as follows:
\begin{isabelle}
\isacommand{corollary}\ Hancl3corollary:\isanewline
\ \ \isakeyword{fixes}\ A::real\ \isakeyword{and}\ \ a\ b\ ::\ "nat\isasymRightarrow int"\isanewline
\ \ \isakeyword{assumes}\ "A\ >\ 1"\ \isakeyword{and}\ a:\ "\isasymforall n.\ a\ n>0"\ \isakeyword{and}\ b:\ "\isasymforall n.\ b\ n>0"\isanewline
\ \ \ \ \isakeyword{and}\ "(\isasymlambda n.\ (a\ n)\ powr(1\ /\ of\_int(2\isacharcircum n)))\ \isasymlonglonglongrightarrow \ A"\isanewline
\ \ \ \ \isakeyword{and}\ "\isasymforall n\ \isasymge \ 6.\ a\ n\ powr(1\ /\ of\_int\ (2\isacharcircum n))\ *\ (1\ +\ 4*(2/3)\isacharcircum n)\ \isasymle \ A\isanewline
\ \ \ \ \ \ \ \ \ \ \ \ \ \ \ \ \ \ \ \ \ \ \ \ \isasymand \ b\ n\ \isasymle \ 2\ powr\ (4/3)\isacharcircum (n-1)"\isanewline
\ \ \isakeyword{shows}\ "(\isasymSum n.\ b\ n\ /\ a\ n)\ \isasymnotin \ \isasymrat "
\end{isabelle}

It is interesting to note that while in Corollary 4.1 (\cite[Corollary 2]{hancl}) we have an assumption that holds ``for all sufficiently large positive integer $n$'', in our formalized version above this assumption is actually specified as ``for all $n \geq 6$''. In our formalized version of Theorem 4.1 (\cite[Theorem 3]{hancl}) above, ``for all sufficiently large $n$'' had been written simply as ``$\forall n \geq s$'' where $s>0$ is some fixed, unknown number.  
Statements of the form ``for all sufficiently large'' can also be encoded in a more abstract fashion using the keyword \isakeyword{eventually}, which does not require an additional variable.

\subsubsection{Infinite Products}

This material requires the notion of an infinite product, which was then (early 2018) not available in Isabelle except in a small development by Manuel Eberl. Infinite sums were available, and because we were only dealing with positive reals, the infinite products could be reduced to infinite sums via logarithms. However, since one objective of this work was to identify and fill gaps in the libraries, we decided to formalize infinite products directly. 
We extended Eberl's development to a comprehensive \isa{Infinite\_Products} library and included it in the next Isabelle release (Isabelle 2018). 

Consider the following technical lemma, where  
we make use of \isa{\isasymProd}: 
\begin{isabelle}
\isacommand{lemma}\ show8:\isanewline
\ \ \isakeyword{fixes}\ d\ ::"nat\isasymRightarrow real\ "\ \isakeyword{and}\ a\ ::"nat\isasymRightarrow int"\ \isakeyword{and}\ s\ k::nat\ \isanewline
\ \ \isakeyword{assumes}\ "A\ >\ 1"\ \isakeyword{and}\ d:\ "\isasymforall n.\ d\ n>\ 1"\ \isakeyword{and}\ "\isasymforall n.\ a\ n>0"\ \isakeyword{and}\ "s>0"\isanewline
\ \ \ \ \isakeyword{and}\ "convergent\_prod\ d"\isanewline
\ \ \ \ \isakeyword{and}\ "\isasymforall n\ \isasymge \ s.\ A\ /\ (a\ n)\ powr(1\ /\ (2::int)\isacharcircum n)\ >\ (\isasymProd j.\ d(n\ +j))"\isanewline
\ \ \isakeyword{shows}\ "\isasymforall n\isasymge s.\ (\isasymProd j.\ d\ (j\ +\ n))\ <\ A\ \isanewline
\ \ \ \ \ \ \ \ \ \ \ \ \ \ \ \ \ \ \ \ \ \ \ /\ (MAX\ j\isasymin \isacharbraceleft s..n\isacharbraceright .\ a\ j\ powr\ (1\ /\ (2::int)\ \isacharcircum \ j))"
\end{isabelle}

This had been originally formulated using \isa{\isasymSum} via the natural logarithm in the first version, so the conclusion instead read 
\begin{isabelle}
\isakeyword{shows}\ "\isasymforall n\isasymge s.\ exp\ (\isasymSum j.\ ln(d\ (j\ +\ n)))\ <\ A\ \isanewline
\ \ \ \ \ \ \ \ \ \ \ \ \ \ \ \ \ \ \ \ \ \ \ /\ (MAX\ j\isasymin \isacharbraceleft s..n\isacharbraceright .\ a\ j\ powr\ (1\ /\ (2::int)\ \isacharcircum \ j))"
\end{isabelle}

By convention, infinite products are defined in two stages \cite[p.\ts241]{bak-complex-analysis}. 
\begin{enumerate}
    \item For nonzero complex numbers, the infinite product $\prod_{k=0}^\infty u_k$ is defined to \emph{converge} if the finite partial products converge to a nonzero limit. If that limit is zero, then the infinite product \emph{diverges} to zero. 
    \item If finitely many of the $u_k$ equal zero and the infinite product of the nonzero terms converges in the sense of (1), then $\prod_{k=0}^\infty u_k$ converges to zero.
\end{enumerate}
One advantage of this approach is that $\prod_{k=0}^\infty u_k = u_0 \prod_{k=1}^\infty u_k$ and similar identities hold.

\section{The transcendence of certain infinite series}
\subsection{Results formalized}
The following results are by Han\v{c}l and Rucki in \cite{hanclrucki}.
 
\begin{theorem}[\hbox{Han\v{c}l and Rucki \cite[Theorem 2.1]{hanclrucki}}]
Let $\delta$ be a positive real number. Let $\{\alpha_n\}_{n=1}^\infty$ and $\{b_n\}_{n=1}^\infty$ be sequences of positive integers such that
$$\limsup_{n \rightarrow \infty}  \frac{\alpha_{n+1}}{(\prod^n_{i=1} \alpha_i)^{2+\delta}}\cdot \frac{1}{b_{n+1}} = \infty $$
and
$$\liminf_{n \rightarrow \infty} \frac{\alpha_{n+1}}{\alpha_n} \cdot \frac{b_n}{b_{n+1}}>1. $$
Then the sum $$\sum^\infty_{n=1} \frac{b_n}{\alpha_n}$$ is transcendental.
\end{theorem}

\begin{theorem}[\hbox{Han\v{c}l and Rucki \cite[Theorem 2.2]{hanclrucki}}] Let $\delta$ and $\epsilon$ be positive real numbers. 
Let $\{\alpha_n\}_{n=1}^\infty$ and $\{b_n\}_{n=1}^\infty$ be sequences of positive integers such that
$$\limsup_{n \rightarrow \infty} \frac{\alpha_{n+1}}{( \prod^n_{i=1} \alpha_i )^{2+ 2/\epsilon +\delta}}\cdot \frac{1}{b_{n+1}} = \infty $$
and for every sufficiently large $n$
$$\sqrt[1+\epsilon]{\frac{\alpha_{n+1}}{b_{n+1}}} \geq \sqrt[1+\epsilon]{ \frac{\alpha_{n}}{b_{n}}}+1.$$
Then the sum $$\sum^\infty_{n=1} \frac{b_n}{\alpha_n}$$ is transcendental.
\end{theorem}

The two theorems above are in fact corollaries of Roth's celebrated result on rational approximations to algebraic numbers from 1955:

\begin{theorem}[Roth \cite{Roth}] Let $a$ be any algebraic number, not rational. If the inequality
$$|a - \frac{p}{q} |< \frac{1}{q^\kappa}$$ has infinitely many solutions in coprime integers $p$ and $q$ where $q>0$, then $\kappa \leq 2$.
\end{theorem}
In particular, the two theorems above by Han\v{c}l and Rucki \cite{hanclrucki} asserting 
the transcendence of the sum of certain series are shown by finding infinitely many such integer solutions 
of the inequality with $a$ the sum of the series in question and for some  $\kappa >2$.

\subsection{On the formalization}
Our full formalization is online \cite{transhanclrucki}.\footnote{The reader may notice that this development too seems to depend on the AFP entry for the Prime Number Theorem \cite{PrimeNumberTheorem}. However we only 
use some technical lemmas from that development. Primes play no role here.}
In this work, too, we had to fill in certain arguments that had been omitted in the original paper, for example to first show that the series in question were summable.

\subsubsection{Implementing Roth's Theorem}
As already mentioned, for the proofs of Theorems 5.1 (\cite[Theorem 2.1]{hanclrucki}) and 5.2 (\cite[Theorem 2.2]{hanclrucki}), 
we applied Theorem 5.3: Roth's theorem on rational approximations to algebraic numbers \cite{Roth}.
The proof of this celebrated theorem is long and elaborate; it has not been formalized in any proof assistant, to our knowledge.
In our formalization we have thus used the statement of Roth's theorem merely as an assumption, instead of formalizing Roth's proof itself beforehand. Roth's theorem was implemented within a \emph{locale} as follows:
\begin{isabelle}
\isacommand{locale}\ RothsTheorem\ =\ \isanewline
\ \ \isakeyword{assumes}\ RothsTheorem:\isanewline
\ \ \ \ \ \ "\isasymforall \isasymxi \ \isasymkappa .\ algebraic\ \isasymxi \ \isasymand \ \isasymxi \ \isasymnotin \ \isasymrat \ \isasymand \ infinite\ \isacharbraceleft (p,q).\ (q::int)>0\ \isasymand \ \isanewline
\ \ \ \ \ \ \ \ \ \ \ \ \ \ coprime\ p\ q\ \isasymand \ \isasymbar \isasymxi \ -\ \ p/q\isasymbar \ <\ 1/q\ powr\ \isasymkappa \isacharbraceright \ \isasymlongrightarrow \ \isasymkappa \ \isasymle \ 2"
\end{isabelle}
A locale collects parameters and assumptions, which it packages as a context in which to work. This locale simply packages the assumption that Roth's theorem is true. It is a safer option than assuming the theorem as an axiom. Theorems 5.1 (\cite[Theorem 2.1]{hanclrucki}) and 5.2 (\cite[Theorem 2.2]{hanclrucki}) were then formalized within this locale:
\begin{isabelle}
\isacommand{theorem}\ (\isakeyword{in}\ RothsTheorem)\ HanclRucki1:\isanewline
\ \ \isakeyword{fixes}\ a\ b\ ::\ "nat\ \isasymRightarrow \ int"\ \isakeyword{and}\ \isasymdelta \ ::\ real\ \isanewline
\ \ \isakeyword{assumes}\ "\isasymforall k.\ a\ k\ >\ 0"\ \isakeyword{and}\ "\isasymforall k.\ b\ k\ >\ 0"\ \isakeyword{and}\ "\isasymdelta \ >\ 0"\isanewline
\ \ \ \ \isakeyword{and}\ "limsup\ (\isasymlambda k.\ a(k+1)\ /\isanewline
\ \ \ \ \ \ \ \ \ \ \ \ \ \ \ \ \ \ \ \ (\isasymProd i=0..k.\ a\ i)\ powr\ (2+\isasymdelta)\ *\ (1\ /\ b(k+1)))\ =\ \isasyminfinity "\isanewline
\ \ \ \ \isakeyword{and}\ "liminf\ (\isasymlambda k.\ a\ (k+1)\ /\ a\ k\ *\ b\ k\ /\ b\ (k+1))\ >\ 1"\isanewline
\ \ \isakeyword{shows}\ "\isasymnot \ algebraic\ (\isasymSum\ k.\ b\ k\ /\ a\ k)"\isanewline
\end{isabelle}

\begin{isabelle}
\isacommand{theorem}\ (\isakeyword{in}\ RothsTheorem)\ HanclRucki2:\ \isanewline
\ \ \isakeyword{fixes}\ a\ b\ ::\ "nat\isasymRightarrow int"\ \isakeyword{and}\ \isasymdelta \ \isasymepsilon \ ::\ real\ \isakeyword{and}\ t\ ::\ nat\isanewline
\ \ \isakeyword{assumes}\ "\isasymforall k.\ a\ k\ >\ 0"\ \isakeyword{and}\ "\isasymforall k.\ b\ k\ >\ 0"\ \isakeyword{and}\ "\isasymdelta \ >\ 0"\ \ "\isasymepsilon \ >\ 0"\isanewline
\ \ \ \ \isakeyword{and}\ "limsup\ (\isasymlambda k.\ a(k+1)\ /\ (\isasymProd i=0..k.\ a\ i)\ powr\ (2\ +\ 2/\isasymepsilon \ +\ \isasymdelta )\isanewline
\ \ \ \ \ \ \ \ \ \ \ \ \ \ \ \ \ \ \ \ \ \ \ \ \ \ \ \ \ \ \ \ \ \ \ \ \ \ \ \ \ \ \ \ \ \ \ \ *\ (1\ /\ b(k+1)))\ =\ \isasyminfinity "\isanewline
\ \ \ \ \isakeyword{and}\ "\isasymforall k\ \isasymge \ t.\ (a\ (k+1)\ /\ b\ (k+1))\ powr\ (1\ /\ (1+\isasymepsilon ))\ \isanewline
\ \ \ \ \ \ \ \ \ \ \ \ \ \ \ \ \ \ \ \ \ \ \ \ \ \ \ \ \ \ \ \ \ \ \isasymge \ (a\ k\ /\ b\ k)\ powr\ (1\ /\ (1+\isasymepsilon ))\ +\ 1"\isanewline
\ \ \isakeyword{shows}\ "\isasymnot \ algebraic\ (\isasymSum\ k.\ b\ k\ /\ a\ k)"
\end{isabelle}

Assuming a key theorem whose proof has not been formalized could be seen as compromising the vision of 
absolute correctness, which demands a bottom-up approach: to formalize the proofs of all prerequisites.
However, ours is a more realistic approach that reflects actual mathematical practice. It still guarantees the correctness of the arguments we formalized, which assume Roth's theorem. Normally, whenever mathematicians make use of a theorem  in their work, they do not prove it themselves all over again but they trust the proof in the literature. 
But when writing a proof with a proof assistant,  one would ideally expect that 
every result used would also be \textit{formally} verified as well. Here, however, Roth's  theorem is used even though its proof hasn't been formalized; we just trust its \textit{informal} proof \cite{Roth}, just like a traditional mathematician would do.

Our approach has the significant advantage of reaching
the formalization of more advanced mathematics faster.
Nevertheless, care must be taken to avoid propagating errors in the literature.
We recommend that a formally unverified theorem should be taken as an assumption within a formalization only if it is a fundamental result that has been checked 
by many mathematicians in the past. Roth's famous theorem can be considered safe in this sense. And all such assumptions must be declared openly \cite[Section 5]{praxis}.

\subsubsection{A small issue in the original proof}

The process of the formalization helped to reveal a slight mistake in one of the original proofs. It puzzled us for some time, although the eventual fix was straightforward. In the proof of Theorem 5.1
\cite[page 534]{hanclrucki} it is claimed that from the assumptions it follows that for each real number $A>1$, there exists a positive integer $k_0$
such that for all $ k > k_0$, 
$$\frac{1}{A}  \cdot  \frac{b_k}{\alpha_k} > \frac{b_{k+1}}{\alpha_{k+1}}.$$
During the formalization process, we noticed that there is a problem with this claim. Consider the counterexample where we set $\alpha_{k+1} = k (\prod^k_{i=1} \alpha_i)^{\lceil 2+ \delta \rceil}$ if $k$ is odd, and
$\alpha_{k+1}= 2 \alpha_k$ otherwise, with $b_k=1$ for all $k$. To resolve this, we had suggested a slightly
modified version of the original statement and a different proof. However, this turned out to be unnecessary; 
the authors suggested to us via email that the problem could be resolved by replacing ``for each real number $A>1$'' with ``there exists a real number $A>1$'' above.
This suggestion repaired the proof, and the rest of our formalization \cite{transhanclrucki} followed the original proof \cite{hanclrucki} without any further problems.

\section{On the de Bruijn factors}
The de Bruijn factor, introduced in 1977 by L. S. van Benthem Jutting \cite{jutting77}, is the ratio of the size of a formalization (in symbols)  to the size of the original mathematical text (in words). 
Freek Wiedijk \cite{freekdebruijn} recommends making this more precise by comparing the number of bytes in the computer encodings (using the  \LaTeX{} source of the mathematics), even compressing both files to ensure independence from such factors as the lengths of identifiers.
Such precision is of questionable value, given the enormous variations in the density of mathematical texts.

Here we only give approximate estimates of the de Bruijn factors for our formalizations, counting
the number of lines in our formalizations and in the respective published informal proofs .
We consider the entire amount of the material (i.e. statements together with their proofs
as well as corollaries together with their proofs) in each formalization work.
Given our formalization \cite{irrationalityerdosstraus} which spans around 1960 lines, we estimate the de Bruijn factor for Theorem 3.1 (\cite[Theorem 2.1]{erdosstraus}),
Corollary 3.1 (\cite[Corollary 2.10]{erdosstraus}) and Theorem 3.2 (\cite[Theorem 3.1]{erdosstraus}) and their respective proofs altogether to be around 25.
Given our formalization \cite{irrationalityhancl} which spans around 1054 lines, we estimate the de Bruijn factor for Theorem 4.1 (\cite[Theorem 3]{hancl}), Corollary 4.1 (\cite[Corollary 2]{hancl}) and their respective proofs altogether to be around 21.
Finally, given our formalization \cite{transhanclrucki} which spans around 990 lines, we estimate the de Bruijn factor for Theorem 5.1 (\cite[Theorem 2.1]{hanclrucki})
and Theorem 5.2 (\cite[Theorem 2.2]{hanclrucki}) and their respective proofs altogether to be around 13.
These de Bruijn factors can be considered high---Wiedijk reports lower factors for some formalizations in HOL~\cite{freekdebruijnweb}---because we had to fill in a considerable amount of intermediate arguments which had not been made explicit in the original proofs, especially in the case of Erdős--Straus~\cite{erdosstraus}.

\section{Conclusion}

We have formalized, in Isabelle/HOL \cite{irrationalityhancl, transhanclrucki,irrationalityerdosstraus}, several results on 
irrationality and transcendence criteria for infinite series from three research papers:
by Erd\H{o}s--Straus (1974) \cite{erdosstraus}, Han\v{c}l (2002)~\cite{hancl} and Han\v{c}l--Rucki (2005) \cite{hanclrucki}. 

We formalized results from mainstream journal papers in their original, unpolished form. Landmarks in the world of formalized mathematics include proving the four-colour theorem in Coq by Georges Gonthier \cite{gonthier}, the Kepler conjecture \cite{hales} in HOL Light and Isabelle by Thomas C. Hales et al. \cite{halesteam} and, more recently, the solution to the Cap Set Problem in Lean by Dahmen, H\"{o}lzl and Lewis \cite{dahmen_et_al}, the formalization of perfectoid spaces in Lean by Buzzard, Commelin and Massot \cite{buzzardperfectoid} and the formalization of Grothendieck's Schemes in Isabelle/HOL by 
Bordg, Paulson and Li \cite{scheme_afp,scheme_arxiv}.
More in the spirit of our work is the formalization in Isabelle/HOL by Gou\"{e}zel of a recent paper by Shchur
in the theory of Gromov-hyperbolic spaces, where a mistake in the original paper was moreover discovered through the formalization process
\cite{gouezel_shchur}.

Summarizing the lessons that we learned through our formalization, we can now revisit the questions posed in the Introduction:
\begin{itemize}
\item The amount of preliminary content in the current Isabelle/HOL libraries was indeed satisfactory.
We had a priori chosen material that we expected would be within reach, and indeed, we didn't run into significant
deficiencies in the library. Some prerequisites that had to be added were a detailed formalization of infinite products, a modification of the ratio test for series and some facts about prime numbers. All these additions were doable within a reasonable amount of time and 
we can expect that they will be used in many other contexts. Our formalizations were thus an incentive to enrich the library.
\item In general we have found Sledgehammer's automation to be of immense help, as hinted by the many occurrences throughout our formal proofs of proof methods such as \isa{metis} and \isa{smt}, which are generated by Sledgehammer.
\item As discussed in Section 6, the formal proofs were significantly longer than the original, which is reflected by the fairly high de Bruijn factors (approximately 25, 21 and 13 for the material  \cite{erdosstraus, hancl, hanclrucki} respectively).
This was certainly not unexpected.
\item We did not discover any major issues in the original proofs. We did however need to fill in some missing steps or calculations, some of which were proofs of summability of infinite series. These omissions contributed to the high de Bruijn factors.
Regarding Erd\H{o}s--Straus~\cite{erdosstraus}, one of the proofs was slightly reformulated, while in other proofs a few inequalities required some minor corrections that did not, however, affect the correctness of the statements or the general proof structure.
In Han\v{c}l--Rucki~\cite{hanclrucki}, a slight mistake was easily fixed thanks to a hint from the authors.
Our detailed, complete formal proofs may even be useful to the readers who wish to study the original proofs.
\end{itemize}
Overall, we found the formalization process to be interesting and educational. We hope that our feedback
may be useful both to experienced proof assistant users and to mathematicians interested in attempting to formalize some research work for the first time.

\paragraph*{Acknowledgements.}
The authors thank the ERC for their support through the ERC Advanced Grant ALEXANDRIA (Project GA 742178). We thank the anonymous reviewers for their useful suggestions that improved our exposition.
We also thank Jaroslav Han\v{c}l and Pavel Rucki for a helpful email  
suggesting  a fix for the slight mistake in their proof \cite[Theorem 2.1]{hanclrucki}, Iosif Pinelis for his useful explanation on MathOverflow regarding an argument in the proof of Theorem 2.2 in the same paper\footnote{\url{https://mathoverflow.net/questions/323069/why-is-this-series-summable}}
and Anthony Bordg for a discussion on the necessity of formalizing advanced mathematics even if the proofs of the prerequisites are not formalized.

\section*{Appendix}
Our formalized version of Theorem 4.1 (\cite[Theorem 3]{hancl}) along with its formalized proof as in \cite{irrationalityhancl}
is presented below:
\begin{isabelle}
\isacommand{theorem}\isamarkupfalse%
\ Hancl{\isadigit{3}}{\isacharcolon}\isanewline
\ \ \isakeyword{fixes}\ d\ {\isacharcolon}{\isacharcolon}\ {\isachardoublequoteopen}nat{\isasymRightarrow}real\ {\isachardoublequoteclose}\ \isakeyword{and}\ a\ b\ {\isacharcolon}{\isacharcolon}{\isachardoublequoteopen}nat{\isasymRightarrow}int{\isachardoublequoteclose}\ \isakeyword{and}\ A{\isacharcolon}{\isacharcolon}real\ \isakeyword{and}\ s{\isacharcolon}{\isacharcolon}nat\isanewline
\ \ \isakeyword{assumes}\ {\isachardoublequoteopen}A\ {\isachargreater}\ {\isadigit{1}}{\isachardoublequoteclose}\ \isakeyword{and}\ d{\isacharcolon}{\isachardoublequoteopen}{\isasymforall}n{\isachardot}\ d\ n{\isachargreater}\ {\isadigit{1}}{\isachardoublequoteclose}\ \isakeyword{and}\ a{\isacharcolon}{\isachardoublequoteopen}{\isasymforall}n{\isachardot}\ a\ n{\isachargreater}{\isadigit{0}}{\isachardoublequoteclose}\isanewline
\ \ \ \ \isakeyword{and}\ b{\isacharcolon}{\isachardoublequoteopen}{\isasymforall}n{\isachardot}\ b\ n\ {\isachargreater}\ {\isadigit{0}}{\isachardoublequoteclose}\ \isakeyword{and}\ {\isachardoublequoteopen}s\ {\isachargreater}\ {\isadigit{0}}{\isachardoublequoteclose}\isanewline
\ \ \ \ \isakeyword{and}\ assu{\isadigit{1}}{\isacharcolon}\ {\isachardoublequoteopen}{\isacharparenleft}{\isacharparenleft}\ {\isasymlambda}\ n{\isachardot}\ a\ n\ powr\ {\isacharparenleft}{\isadigit{1}}{\isacharslash}{\isacharparenleft}{\isadigit{2}}{\isacharcolon}{\isacharcolon}int{\isacharparenright}{\isacharcircum}n{\isacharparenright}{\isacharparenright}{\isasymlonglongrightarrow}\ A{\isacharparenright}\ sequentially{\isachardoublequoteclose}\isanewline
\ \ \ \ \isakeyword{and}\ assu{\isadigit{2}}{\isacharcolon}\ {\isachardoublequoteopen}{\isasymforall}n\ {\isasymge}\ s{\isachardot}\ A\ {\isacharslash}\ a\ n\ powr\ {\isacharparenleft}{\isadigit{1}}{\isacharslash}{\isacharparenleft}{\isadigit{2}}{\isacharcolon}{\isacharcolon}int{\isacharparenright}{\isacharcircum}n{\isacharparenright}{\isachargreater}\ {\isacharparenleft}{\isasymProd}j{\isachardot}\ d\ {\isacharparenleft}n\ {\isacharplus}\ j{\isacharparenright}{\isacharparenright}{\isachardoublequoteclose}\isanewline
\ \ \ \ \isakeyword{and}\ assu{\isadigit{3}}{\isacharcolon}\ {\isachardoublequoteopen}LIM\ n\ sequentially{\isachardot}\ d\ n\ {\isacharcircum}\ {\isadigit{2}}\ {\isacharcircum}\ n\ {\isacharslash}\ b\ n\ {\isacharcolon}{\isachargreater}\ at{\isacharunderscore}top{\isachardoublequoteclose}\isanewline
\ \ \ \ \isakeyword{and}\ {\isachardoublequoteopen}convergent{\isacharunderscore}prod\ d{\isachardoublequoteclose}\ \isanewline
\ \ \isakeyword{shows}\ {\isachardoublequoteopen}{\isacharparenleft}{\isasymSum}\ n{\isachardot}\ b\ n\ {\isacharslash}\ a\ n{\isacharparenright}\ {\isasymnotin}\ {\isasymrat}{\isachardoublequoteclose}\ \isanewline
\isacommand{proof}\isamarkupfalse%
{\isacharparenleft}rule\ ccontr{\isacharparenright}\isanewline
\ \ \isacommand{assume}\isamarkupfalse%
\ asm{\isacharcolon}{\isachardoublequoteopen}{\isasymnot}\ {\isacharparenleft}{\isacharparenleft}{\isasymSum}n{\isachardot}\ b\ n\ {\isacharslash}\ a\ n\ {\isacharparenright}\ {\isasymnotin}\ {\isasymrat}{\isacharparenright}{\isachardoublequoteclose}\isanewline
\ \ \isacommand{have}\isamarkupfalse%
\ ab{\isacharunderscore}sum{\isacharcolon}{\isachardoublequoteopen}summable\ {\isacharparenleft}{\isasymlambda}j{\isachardot}\ b\ j\ {\isacharslash}\ a\ j{\isacharparenright}{\isachardoublequoteclose}\isanewline
\ \ \ \ \isacommand{using}\isamarkupfalse%
\ issummable{\isacharbrackleft}OF\ {\isacartoucheopen}A{\isachargreater}{\isadigit{1}}{\isacartoucheclose}\ d\ a\ b\ assu{\isadigit{1}}\ assu{\isadigit{3}}\ {\isacartoucheopen}convergent{\isacharunderscore}prod\ d{\isacartoucheclose}{\isacharbrackright}\ \isacommand{{\isachardot}}\isamarkupfalse%
\isanewline
\ \ \isacommand{obtain}\isamarkupfalse%
\ p\ q\ {\isacharcolon}{\isacharcolon}int\ \isakeyword{where}\ {\isachardoublequoteopen}q{\isachargreater}{\isadigit{0}}{\isachardoublequoteclose}\ \isanewline
\ \ \ \ \ \ \ \ \ \ \ \ \ \ \ \ \ \ \ \ \ \ \ \ \ \ \isakeyword{and}\ pq{\isacharunderscore}def{\isacharcolon}{\isachardoublequoteopen}{\isacharparenleft}{\isasymlambda}\ n{\isachardot}\ b\ {\isacharparenleft}n{\isacharplus}{\isadigit{1}}{\isacharparenright}\ {\isacharslash}\ a\ {\isacharparenleft}n{\isacharplus}{\isadigit{1}}{\isacharparenright}{\isacharparenright}\ sums\ {\isacharparenleft}p{\isacharslash}q{\isacharparenright}{\isachardoublequoteclose}\isanewline
\ \ \isacommand{proof}\isamarkupfalse%
\ {\isacharminus}\isanewline
\ \ \ \ \isacommand{from}\isamarkupfalse%
\ asm\ \isacommand{have}\isamarkupfalse%
\ {\isachardoublequoteopen}{\isacharparenleft}{\isasymSum}\ n{\isachardot}\ b\ n\ {\isacharslash}\ a\ n{\isacharparenright}\ {\isasymin}\ {\isasymrat}{\isachardoublequoteclose}\ \isacommand{by}\isamarkupfalse%
\ auto\isanewline
\ \ \ \ \isacommand{then}\isamarkupfalse%
\ \isacommand{have}\isamarkupfalse%
\ {\isachardoublequoteopen}{\isacharparenleft}{\isasymSum}\ n{\isachardot}\ b\ {\isacharparenleft}n{\isacharplus}{\isadigit{1}}{\isacharparenright}\ {\isacharslash}\ a\ {\isacharparenleft}n{\isacharplus}{\isadigit{1}}{\isacharparenright}{\isacharparenright}\ {\isasymin}\ {\isasymrat}{\isachardoublequoteclose}\ \isanewline
\ \ \ \ \ \ \isacommand{by}\isamarkupfalse%
\ {\isacharparenleft}subst\ suminf{\isacharunderscore}minus{\isacharunderscore}initial{\isacharunderscore}segment{\isacharbrackleft}OF\ ab{\isacharunderscore}sum{\isacharcomma}of\ {\isadigit{1}}{\isacharbrackright}{\isacharparenright}\ auto\isanewline
\ \ \ \ \isacommand{then}\isamarkupfalse%
\ \isacommand{obtain}\isamarkupfalse%
\ p{\isacharprime}\ q{\isacharprime}\ {\isacharcolon}{\isacharcolon}int\ \isakeyword{where}\ {\isachardoublequoteopen}q{\isacharprime}{\isasymnoteq}{\isadigit{0}}{\isachardoublequoteclose}\ \isanewline
\ \ \ \ \ \ \ \ \isakeyword{and}\ pq{\isacharunderscore}def{\isacharcolon}{\isachardoublequoteopen}{\isacharparenleft}{\isasymlambda}\ n{\isachardot}\ b\ {\isacharparenleft}n{\isacharplus}{\isadigit{1}}{\isacharparenright}\ {\isacharslash}\ a\ {\isacharparenleft}n{\isacharplus}{\isadigit{1}}{\isacharparenright}\ {\isacharparenright}\ sums\ {\isacharparenleft}p{\isacharprime}{\isacharslash}q{\isacharprime}{\isacharparenright}{\isachardoublequoteclose}\isanewline
\ \ \ \ \ \ \isacommand{unfolding}\isamarkupfalse%
\ Rats{\isacharunderscore}eq{\isacharunderscore}int{\isacharunderscore}div{\isacharunderscore}int\isanewline
\ \ \ \ \ \ \isacommand{using}\isamarkupfalse%
\ summable{\isacharunderscore}ignore{\isacharunderscore}initial{\isacharunderscore}segment{\isacharbrackleft}OF\ ab{\isacharunderscore}sum{\isacharcomma}of\ {\isadigit{1}}{\isacharcomma}THEN\ summable{\isacharunderscore}sums{\isacharbrackright}\isanewline
\ \ \ \ \ \ \isacommand{by}\isamarkupfalse%
\ force\isanewline
\ \ \ \ \isacommand{define}\isamarkupfalse%
\ p\ q\ \isakeyword{where}\ {\isachardoublequoteopen}p{\isacharequal}{\isacharparenleft}if\ q{\isacharprime}{\isacharless}{\isadigit{0}}\ then\ {\isacharminus}\ p{\isacharprime}\ else\ p{\isacharprime}{\isacharparenright}{\isachardoublequoteclose}\ \isanewline
\ \ \ \ \ \ \ \ \ \ \ \ \ \ \ \ \ \isakeyword{and}\ {\isachardoublequoteopen}q{\isacharequal}{\isacharparenleft}if\ q{\isacharprime}{\isacharless}{\isadigit{0}}\ then\ {\isacharminus}\ q{\isacharprime}\ else\ q{\isacharprime}{\isacharparenright}{\isachardoublequoteclose}\isanewline
\ \ \ \ \isacommand{have}\isamarkupfalse%
\ {\isachardoublequoteopen}p{\isacharprime}{\isacharslash}q{\isacharprime}{\isacharequal}p{\isacharslash}q{\isachardoublequoteclose}\ \isakeyword{and}\ {\isachardoublequoteopen}q{\isachargreater}{\isadigit{0}}{\isachardoublequoteclose}\ \isanewline
\ \ \ \ \ \ \isacommand{using}\isamarkupfalse%
\ {\isacartoucheopen}q{\isacharprime}{\isasymnoteq}{\isadigit{0}}{\isacartoucheclose}\ \isacommand{unfolding}\isamarkupfalse%
\ p{\isacharunderscore}def\ q{\isacharunderscore}def\ \isacommand{by}\isamarkupfalse%
\ auto\isanewline
\ \ \ \ \isacommand{then}\isamarkupfalse%
\ \isacommand{show}\isamarkupfalse%
\ {\isacharquery}thesis\ \isacommand{using}\isamarkupfalse%
\ that{\isacharbrackleft}of\ q\ p{\isacharbrackright}\ pq{\isacharunderscore}def\ \isacommand{by}\isamarkupfalse%
\ auto\isanewline
\ \ \isacommand{qed}\isamarkupfalse%
\isanewline
\isanewline
\ \ \isacommand{define}\isamarkupfalse%
\ ALPHA\ \isakeyword{where}\ \isanewline
\ \ \ \ {\isachardoublequoteopen}ALPHA\ {\isacharequal}\ {\isacharparenleft}{\isasymlambda}n{\isachardot}\ q\ {\isacharasterisk}\ {\isacharparenleft}{\isasymProd}j{\isacharequal}{\isadigit{1}}{\isachardot}{\isachardot}n{\isachardot}\ a\ j{\isacharparenright}\ {\isacharasterisk}\ {\isacharparenleft}{\isasymSum}j{\isachardot}\ b\ {\isacharparenleft}j{\isacharplus}n{\isacharplus}{\isadigit{1}}{\isacharparenright}\ {\isacharslash}\ a\ {\isacharparenleft}j{\isacharplus}n{\isacharplus}{\isadigit{1}}{\isacharparenright}{\isacharparenright}{\isacharparenright}{\isachardoublequoteclose}\isanewline
\ \ \isacommand{have}\isamarkupfalse%
\ {\isachardoublequoteopen}ALPHA\ n\ {\isasymge}\ {\isadigit{1}}{\isachardoublequoteclose}\ \isakeyword{for}\ n\ \isanewline
\ \ \isacommand{proof}\isamarkupfalse%
\ {\isacharminus}\isanewline
\ \ \ \ \isacommand{have}\isamarkupfalse%
\ {\isachardoublequoteopen}{\isacharparenleft}{\isasymSum}\ n{\isachardot}\ b\ {\isacharparenleft}n{\isacharplus}{\isadigit{1}}{\isacharparenright}\ {\isacharslash}\ a\ {\isacharparenleft}n{\isacharplus}{\isadigit{1}}{\isacharparenright}{\isacharparenright}\ {\isachargreater}\ {\isadigit{0}}{\isachardoublequoteclose}\isanewline
\ \ \ \ \ \ \isacommand{apply}\isamarkupfalse%
\ {\isacharparenleft}rule\ suminf{\isacharunderscore}pos{\isacharparenright}\ \isanewline
\ \ \ \ \ \ \isacommand{using}\isamarkupfalse%
\ summable{\isacharunderscore}ignore{\isacharunderscore}initial{\isacharunderscore}segment{\isacharbrackleft}OF\ ab{\isacharunderscore}sum{\isacharcomma}of\ {\isadigit{1}}{\isacharbrackright}\ a\ b\ \isacommand{by}\isamarkupfalse%
\ auto\isanewline
\ \ \ \ \isacommand{then}\isamarkupfalse%
\ \isacommand{have}\isamarkupfalse%
\ {\isachardoublequoteopen}p{\isacharslash}q\ {\isachargreater}\ {\isadigit{0}}{\isachardoublequoteclose}\ \isacommand{using}\isamarkupfalse%
\ sums{\isacharunderscore}unique{\isacharbrackleft}OF\ pq{\isacharunderscore}def{\isacharcomma}symmetric{\isacharbrackright}\ \isacommand{by}\isamarkupfalse%
\ auto\isanewline
\ \ \ \ \isacommand{then}\isamarkupfalse%
\ \isacommand{have}\isamarkupfalse%
\ {\isachardoublequoteopen}q{\isasymge}{\isadigit{1}}{\isachardoublequoteclose}\ \isakeyword{and}\ {\isachardoublequoteopen}p{\isasymge}{\isadigit{1}}{\isachardoublequoteclose}\ \isacommand{using}\isamarkupfalse%
\ {\isacartoucheopen}q{\isachargreater}{\isadigit{0}}{\isacartoucheclose}\ \isacommand{by}\isamarkupfalse%
\ {\isacharparenleft}auto\ simp\ add{\isacharcolon}divide{\isacharunderscore}simps{\isacharparenright}\isanewline
\ \ \ \ \isacommand{moreover}\isamarkupfalse%
\ \isacommand{have}\isamarkupfalse%
\ {\isachardoublequoteopen}{\isasymforall}n{\isachardot}\ {\isadigit{1}}\ {\isasymle}\ a\ n{\isachardoublequoteclose}\ \isakeyword{and}\ {\isachardoublequoteopen}{\isasymforall}n{\isachardot}\ {\isadigit{1}}\ {\isasymle}\ b\ n{\isachardoublequoteclose}\ \isanewline
\ \ \ \ \ \ \isacommand{using}\isamarkupfalse%
\ a\ b\ \isacommand{by}\isamarkupfalse%
\ {\isacharparenleft}auto\ simp\ add{\isacharcolon}\ int{\isacharunderscore}one{\isacharunderscore}le{\isacharunderscore}iff{\isacharunderscore}zero{\isacharunderscore}less{\isacharparenright}\isanewline
\ \ \ \ \isacommand{ultimately}\isamarkupfalse%
\ \isacommand{show}\isamarkupfalse%
\ {\isacharquery}thesis\ \isacommand{unfolding}\isamarkupfalse%
\ ALPHA{\isacharunderscore}def\isanewline
\ \ \ \ \ \ \isacommand{using}\isamarkupfalse%
\ show{\isadigit{7}}{\isacharbrackleft}OF\ {\isacharunderscore}\ {\isacharunderscore}\ {\isacharunderscore}\ {\isacharunderscore}\ pq{\isacharunderscore}def{\isacharbrackright}\ \isacommand{by}\isamarkupfalse%
\ auto\isanewline
\ \ \isacommand{qed}\isamarkupfalse%
\isanewline
\ \ \isacommand{moreover}\isamarkupfalse%
\ \isacommand{have}\isamarkupfalse%
\ {\isachardoublequoteopen}{\isasymexists}n{\isachardot}\ ALPHA\ n\ {\isacharless}\ {\isadigit{1}}{\isachardoublequoteclose}\ \isacommand{unfolding}\isamarkupfalse%
\ ALPHA{\isacharunderscore}def\isanewline
\ \ \isacommand{proof}\isamarkupfalse%
\ {\isacharparenleft}rule\ lasttoshow{\isacharbrackleft}OF\ d\ a\ {\isacartoucheopen}s{\isachargreater}{\isadigit{0}}{\isacartoucheclose}\ {\isacartoucheopen}q{\isachargreater}{\isadigit{0}}{\isacartoucheclose}\ {\isacartoucheopen}A{\isachargreater}{\isadigit{1}}{\isacartoucheclose}\ b\ {\isacharunderscore}\ assu{\isadigit{3}}{\isacharbrackright}{\isacharparenright}\isanewline
\ \ \ \ \isacommand{show}\isamarkupfalse%
\ {\isachardoublequoteopen}{\isasymforall}\isactrlsub F\ n\ in\ sequentially{\isachardot}\ {\isacharparenleft}{\isasymSum}j{\isachardot}\ b\ {\isacharparenleft}n\ {\isacharplus}\ j{\isacharparenright}\ {\isacharslash}\ a{\isacharparenleft}n\ {\isacharplus}\ j{\isacharparenright}{\isacharparenright}\ \isanewline
\ \ \ \ \ \ \ \ \ \ \ \ \ \ \ \ \ \ \ \ \ \ \ \ \ \ \ \ \ \ \ \ \ \ \ \ \ \ \ \ \ \ \ \ \ \ \ \ \ \ \ {\isasymle}\ {\isadigit{2}}\ {\isacharasterisk}\ b\ n\ {\isacharslash}\ a\ n{\isachardoublequoteclose}\isanewline
\ \ \ \ \ \ \isacommand{using}\isamarkupfalse%
\ show{\isadigit{5}}{\isacharbrackleft}OF\ {\isacartoucheopen}A{\isachargreater}{\isadigit{1}}{\isacartoucheclose}\ d\ a\ b\ assu{\isadigit{1}}\ assu{\isadigit{3}}\ {\isacartoucheopen}convergent{\isacharunderscore}prod\ d{\isacartoucheclose}{\isacharbrackright}\ \isacommand{by}\isamarkupfalse%
\ simp\isanewline
\ \ \ \ \isacommand{show}\isamarkupfalse%
\ {\isachardoublequoteopen}{\isasymforall}m{\isasymge}s{\isachardot}\ {\isasymexists}n{\isasymge}m{\isachardot}\ d\ {\isacharparenleft}n\ {\isacharplus}\ {\isadigit{1}}{\isacharparenright}\isanewline
\ \ \ \ \ \ \ \ \ \ \ \ \ \ \ \ \ \ \ \ \ \ {\isacharasterisk}\ {\isacharparenleft}MAX\ j{\isasymin}{\isacharbraceleft}s{\isachardot}{\isachardot}n{\isacharbraceright}{\isachardot}\ a\ j\ powr\ {\isacharparenleft}{\isadigit{1}}\ {\isacharslash}\ {\isacharparenleft}{\isadigit{2}}{\isacharcolon}{\isacharcolon}int{\isacharparenright}\ {\isacharcircum}\ j{\isacharparenright}{\isacharparenright}\isanewline
\ \ \ \ \ \ \ \ \ \ \ \ \ \ \ \ \ \ \ \ \ \ \ \ \ \ \ \ \ \ \ {\isasymle}\ a\ {\isacharparenleft}n\ {\isacharplus}\ {\isadigit{1}}{\isacharparenright}\ powr\ {\isacharparenleft}{\isadigit{1}}\ {\isacharslash}\ {\isacharparenleft}{\isadigit{2}}{\isacharcolon}{\isacharcolon}int{\isacharparenright}\ {\isacharcircum}\ {\isacharparenleft}n\ {\isacharplus}\ {\isadigit{1}}{\isacharparenright}{\isacharparenright}{\isachardoublequoteclose}\isanewline
\ \ \ \ \ \ \isacommand{apply}\isamarkupfalse%
\ {\isacharparenleft}rule\ show{\isadigit{9}}{\isacharbrackleft}OF\ {\isacartoucheopen}A{\isachargreater}{\isadigit{1}}{\isacartoucheclose}\ d\ a\ {\isacartoucheopen}s{\isachargreater}{\isadigit{0}}{\isacartoucheclose}\ assu{\isadigit{1}}\ {\isacartoucheopen}convergent{\isacharunderscore}prod\ d{\isacartoucheclose}{\isacharbrackright}{\isacharparenright}\isanewline
\ \ \ \ \ \ \isacommand{using}\isamarkupfalse%
\ show{\isadigit{8}}{\isacharbrackleft}OF\ {\isacartoucheopen}A{\isachargreater}{\isadigit{1}}{\isacartoucheclose}\ d\ a\ {\isacartoucheopen}s{\isachargreater}{\isadigit{0}}{\isacartoucheclose}\ {\isacartoucheopen}convergent{\isacharunderscore}prod\ d{\isacartoucheclose}\ assu{\isadigit{2}}{\isacharbrackright}\ \isanewline
\ \ \ \ \ \ \isacommand{by}\isamarkupfalse%
\ {\isacharparenleft}simp\ add{\isacharcolon}algebra{\isacharunderscore}simps{\isacharparenright}\isanewline
\ \ \isacommand{qed}\isamarkupfalse%
\isanewline
\ \ \isacommand{ultimately}\isamarkupfalse%
\ \isacommand{show}\isamarkupfalse%
\ False\ \isacommand{using}\isamarkupfalse%
\ not{\isacharunderscore}le\ \isacommand{by}\isamarkupfalse%
\ blast\isanewline
\isacommand{qed}
\end{isabelle}

\end{document}